\documentclass{aastex6}

\begin{document}

\title{Exploring the Energy Sources Powering the Light Curve of the Type Ibn Supernova PS15dpn and the
Mass-Loss History of the SN Progenitor}
\author{Shan-Qin Wang\altaffilmark{1}, and Long Li\altaffilmark{2,3}}

\begin{abstract}

PS15dpn is a luminous rapidly rising Type Ibn supernova (SN) discovered
by Pan-STARRS1 (PS1). Previous study showed that its bolometric
light curve (LC) cannot be explained by the $^{56}$Ni model.
In this paper, we used the $^{56}$Ni model, the magnetar model, the
circumstellar interaction (CSI) model, and the CSI plus
$^{56}$Ni model to fit the bolometric LC of PS15dpn. We found that
the $^{56}$Ni model can fit the bolometric LC but the parameters
are unrealistic, and that the magnetar model, the CSI model, and the CSI
plus $^{56}$Ni model can match the data with reasonable parameters. 
Considering the facts that the emission lines indicative of the
interaction between the ejecta and the CSM have been confirmed, and that
the SNe produced by the explosions of massive stars can synthesize moderate
amount of $^{56}$Ni, we suggest that the CSI plus $^{56}$Ni model is the
most promising one. 
Assuming that the CSM is a shell (wind), the masses of the ejecta,
the CSM, and the $^{56}$Ni are $15.79_{-4.77}^{+5.44}$\,M$_\odot$
($14.18_{-1.64}^{+1.81}$\,M$_\odot$), $0.84_{-0.10}^{+0.13}$\,M$_\odot$
($0.88_{-0.12}^{+0.11}$\,M$_\odot$), and $0.32_{-0.11}^{+0.11}$\,M$_\odot$
($0.16_{-0.08}^{+0.13}$\,M$_\odot$), respectively. The inferred
ejecta masses are consistent with the scenario that the progenitors
of SNe Ibn are massive Wolf-Rayet stars. Adopting the shell CSM scenario,
the shell might be expelled by an eruption of the progenitor
just $\sim$ 17--167 days prior to the SN explosion;
for the wind scenario, the inferred mass-loss rate of the wind is
$\sim 8.0$ M$_\odot$\,yr$^{-1}$, indicating that the wind is a
``super-wind" having extreme high mass-loss rate. 

\end{abstract}


\affil{\altaffilmark{1}Guangxi Key Laboratory for Relativistic Astrophysics,
School of Physical Science and Technology, Guangxi University, Nanning 530004,
China; shanqinwang@gxu.edu.cn}
\affil{\altaffilmark{2}School of Astronomy and Space Science, Nanjing
University, Nanjing 210093, China}
\affil{\altaffilmark{3}Key Laboratory of Modern Astronomy and Astrophysics
(Nanjing University), Ministry of Education, China}

\section{Introduction}

\label{sec:Intro}

The interactions of supernova (SN) ejecta and the circumstellar medium (CSM)
from the pre-SN outbursts {or winds can produce} narrow and
intermediate-width emission lines and enhance the luminosities of SNe.
The SNe showing interaction evidence are therefore called ``interacting SNe"
\citep{Smith2017}. According to the types of the emission lines and the physical
nature, interacting SNe can be divided into at least three classes: type IIn
\citep{Sch1990,Fil1997} SNe emitting H$\alpha$ emission lines, type Ibn SNe
\citep{Pas2016,Hoss2017} with spectra showing He emission lines, and type Ia-CSM
\citep{Sil2013} whose spectra also show H$\alpha$ emission lines.

To date, at least 38 SNe Ibn have been confirmed.\footnote{https://sne.space/}
The shapes of light curves (LCs) of SNe Ibn are rather homogeneous
\citep{Hoss2017}. Some SNe Ibn (e.g., SN 2011hw
\citep{Smith2012,Pas2015a}, OGLE-2012-SN-006 \citep{Pas2015b},
and OGLE-2014-SN-131 \citep{Kara2017}) having very slow-declined LCs,
while most SNe Ibn have very fast-declined LCs.

According to the sample collected by \cite{Hoss2017}, the rise time
of about half of SNe Ibn are smaller than 10 days. Furthermore, some
SNe Ibn have extreme short rise time ($\lesssim 5$ days, see Table 4 of
\citealt{Hoss2017}), resembling that of Fast Blue Optical Transients
(FBOTs, see, e.g., \citealt{Drout2014, Arc2016,Tanaka2016,Whit2017,
Pren2018,Pur2018,Per2019}). \cite{Fox2019} compared the optical
and spectral properties of AT~2018cow which is a nearby FBOT with those
of SNe Ibn and IIn, finding some similarities and suggesting that
AT~2018cow and some other FBOTs might be SNe Ibn.

Determining the energy sources powering the LCs of SNe Ibn is
an important and difficult task. For example, \cite{Tominaga2008} suggested
that the LC of SN~2006jc which is a well-studied SN Ibn can be explained by
the $^{56}$Ni model; \cite{Pas2008} got the best-fitting LC by taking into
account the contributions from the He envelope recombination as well as the
$^{56}$Ni cascade decay; \cite{Chu2009} demonstrated that the circumstellar
interaction (CSI) model models with and without $^{56}$Ni can account for the
bolometric LC of SN~2006jc.

Investigating the sites of the SNe Ibn would provide useful information
for determining their progenitors.
While almost all SNe Ibn were discovered in star-forming regions of
the host galaxies and their progenitors have been thought to be
very massive {hydrogen-poor} Wolf-Rayet stars \citep{Pas2016},
one SN Ibn (PS1-12sk) was found on {the outskirts of}
a bright elliptical galaxy where
the star forming is inactive and its progenitor is elusive \citep{Hoss2019}.

In this paper, we study a luminous rapidly rising SN, PS15dpn, {that was}
discovered by Pan-STARRS1 (PS1) in the area of the gravitational-wave GW151226
\citep{Abbott2016} which was produced by a merger of a black hole-black hole (BH-BH)
binary {at} a luminosity distance ($D_{\rm L}$) of $440^{+190}_{-180}$ Mpc.
The redshift ($z$) of PS15dpn is $0.1747\pm0.0001$ \citep{Smartt2016},
corresponding to a luminosity distance of $854$~Mpc \citep{Smartt2016}.

\cite{Smartt2016} collected eight spectra of PS15dpn obtained by
GMOS, PESSTO, and SNIFS (see their Figure 4). The early-epoch spectra of
PS15dpn are very blue \cite{Chambers2016,Smartt2016}, indicating that the
SN was rather hot in its early evolution. The He {\sc i} {$\lambda$}5016,
5875, 7065~{\AA} emission found in the GMOS spectrum \citep{Palazzi2016}
at +26 days provide the key evidence for classifying PS15dpn as a type Ibn
SN. According to the right panel of Figure 4 of \cite{Smartt2016}, the
spectrum of PS15dpn 4 days before its peak luminosity resembles the spectrum
of SN~2010al when its luminosity peaked, while the spectrum of PS15dpn +26 days
post-peak is similar to that of SN~2006jc +31 days post-peak. All the spectra
of these SNe Ibn show clear He {\sc i} and He {\sc ii} emission lines.

Using a third order polynomial fit, \citep{Smartt2016} found that
the rise time of PS15dpn is $\sim 11.6\pm2.45$ days. This result indicates that
the rest-frame rise time of PS15dpn is $\sim 9.87\pm2.09$ days. The post-peak
decline rates of $g-$, $r-$, $i-$, $z-$, and $y-$band LCs are
respectively 0.08, 0.08, 0.06, 0.06, and 0.05 mag day$^{-1}$ (see Fig. 2 of
\citealt{Smartt2016}), comparable to that of $R-$band LC of SN~2002ao
($0.082\pm0.004$ mag day$^{-1}$), SN~2010al ($0.064\pm0.009$ mag day$^{-1}$), and
iPTF13beo ($0.071\pm0.007$ mag day$^{-1}$) \citep{Hoss2017}, and smaller (shallower)
than that of a major fraction of SNe Ibn listed in Table 4 of \cite{Hoss2017}.

The peak magnitudes of $g-$, $r-$, $i-$, $z-$, and $y-$band LCs are
$\sim$19.56, 19.72, 19.81, 19.89, and 20.05 mag, respectively (see Fig. 2 of
\cite{Smartt2016}). Adopting the distance modulus
($\mu=m-M=5~{\rm log}_{10}{(D_{\rm L}/10~{\rm pc})}=39.64$ mag), the peak
absolute magnitudes of $g-$, $r-$, $i-$, $z-$, and $y-$band LCs are
$\sim-$20.08, $-$19.93, $-$19.83, $-$19.76, and $-$19.60 mag, respectively.
The peak magnitudes of multi-band LCs are comparable to that of the
brighter SNe Ibn in the sample of \cite{Hoss2017}, but dimmer than the
threshold ($-21$ mag, \citealt{Gal2012}) of the peak magnitude of a
superluminous SN (SLSN).\footnote{To date, the unique SLSN Ibn is
ASASSN-14ms whose peak luminosity is $\sim 1.7 \times10^{44}$ erg s$^{-1}$
\citep{Vall2018}.}

Based on the multi-band photometry, \cite{Smartt2016} derived the
bolometric LC (see Fig. 2 of \cite{Smartt2016}) of PS15dpn from which
we found that the peak luminosity of the bolometric LC is
$\sim 4\times10^{43}$ erg s$^{-1}$, and that the bolometric LC after
the peak resembles that of SN~2006jc but showed a flattening at the
late epoch.

The facts that the luminosity distance ($854$~Mpc) of PS15dpn is
inconsistent with the estimated luminosity distance ($440^{+190}_{-180}$ Mpc)
of the BH-BH binary producing GW151226 and that BH-BH mergers cannot
produce SNe {exclude} the possibility that PS15dpn is the
electromagnetic counterpart of GW151226.

Nevertheless, PS15dpn itself is an interesting object since: (1)
it is a rapidly rising ($t_{\rm peak}-t_{\rm explosion}\sim 9.9\pm2.1$ days)
SN and \citet{Smartt2016} demonstrated that it cannot be explained by the $^{56}$Ni model;
(2) It is widely believed that the interactions between the
ejecta of interacting SNe (including SNe IIn and Ibn) and their CSM would
power their unusual LCs; studying the properties of the ejecta and
CSM would provide important information, e.g., the masses
of the ejecta and the CSM, the mass-loss rate and mass-loss
history of the SN progenitors, and so on (see \citealt{Smith2014}
for a review and references therein).

The aim of this paper is exploring the possible energy sources of
PS15dpn, the properties of the CSM surrounding this SN, as well as
the mass-loss history of its progenitor. In Section \ref{sec:fit},
we use {some} models to fit the bolometric LC of PS15dpn and derive
the best-fitting parameters. Our discussion and conclusions can be found in
Sections \ref{sec:dis} and \ref{sec:con}, respectively.

\section{Modeling the Bolometric LC of PS15dpn}
\label{sec:fit}

In this section, we use several models to fit the bolometric LC
of PS15dpn. To get best-fitting parameters, we adopt the
Markov Chain Monte Carlo (MCMC) by using the \texttt{emcee} Python package
\citep{Foreman-Mackey2013}. We employ 20 walkers, each walker runs 40,000
steps. We assume uniform priors for all parameters (see the tables below).
After doing MCMC, the best-fitting parameters are yielded by measuring
the medians of the posterior samples. The uncertainties is $1\sigma$ confidence,
corresponding to 16th and 84th percentiles of the posterior samples.

The functional forms and the description details of the models we use
can be found in \citet{Arn1982}, \citet{Cha2012}, \citet{Wang2015b},
and \citet{Liu2018}.

\subsection{The $^{56}$Ni Model}

As mentioned above, \citet{Smartt2016} has shown that the $^{56}$Ni model
cannot fit the bolometric LC of PS15dpn. The best-fitting parameters
derived by \citet{Smartt2016} are $E_{\mathrm{exp}}=5\times10^{51}$~erg,
$M_{\mathrm{ej}}=1.9$\,M$_\odot$, and $M_{\mathrm{Ni}}=1.7$\,M$_\odot$.
Using the equation $E_{\mathrm{exp}}=3/10 M_{\mathrm{ej}}v_{\mathrm{sc}}^2$,
one can derive that the value of the scale velocity of the ejecta $v_{\mathrm{sc}}$
is $\sim 2.1 \times 10^9$~cm s$^{-1}$.

Here, we employ the $^{56}$Ni model \citet{Arn1982,Cha2012} to model the LC,
testing the conclusion obtained by \citet{Smartt2016}.
The free parameters of the $^{56}$Ni model adopted here are
\footnote{Throughout this paper, the optical opacity $\kappa$ is fixed to be
0.1 cm$^2$ g$^{-1}$ (e.g.,\citealt{Whe2015}).}
the ejecta mass $M_{\mathrm{ej}}$,
the scale velocity of the ejecta $v_{\mathrm{sc}}$,
the $^{56}$Ni mass $M_{\mathrm{Ni}}$, the gamma-ray opacity of $^{56}$Ni decay
photons $\kappa_{\gamma,\mathrm{Ni}}$, and the moment of explosion $t_\mathrm{expl}$.

\begin{figure}[tbph]
\begin{center}
\includegraphics[width=0.45\textwidth,angle=0]{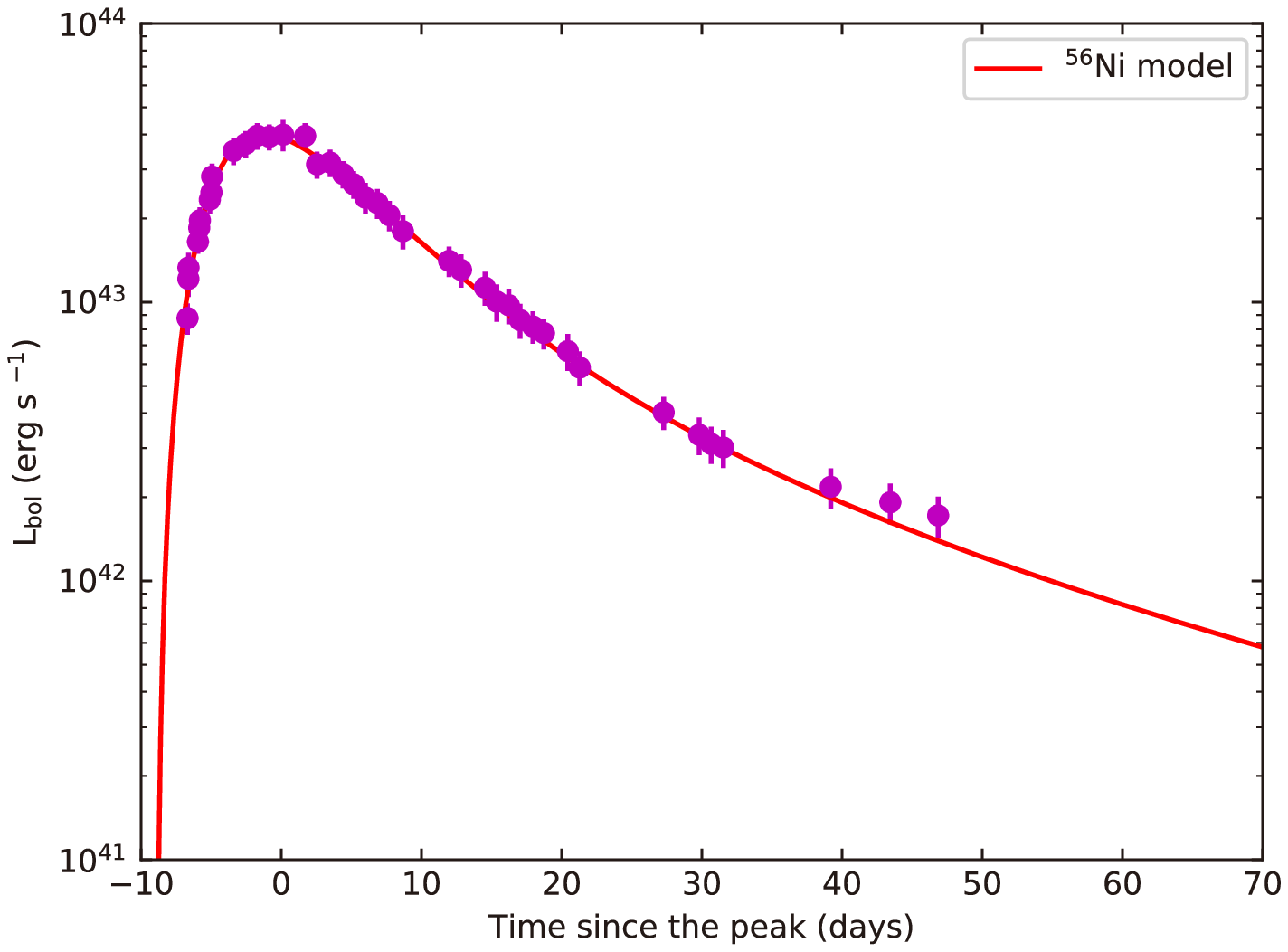}
\end{center}
\caption{The bolometric LC reproduced by the $^{56}$Ni model.
Data are taken from \citet{Smartt2016}.
The abscissa represents time since the peak in the rest frame.}
\label{fig:ni}
\end{figure}

\begin{figure}[tbph]
\begin{center}
\includegraphics[width=1.0\textwidth,angle=0]{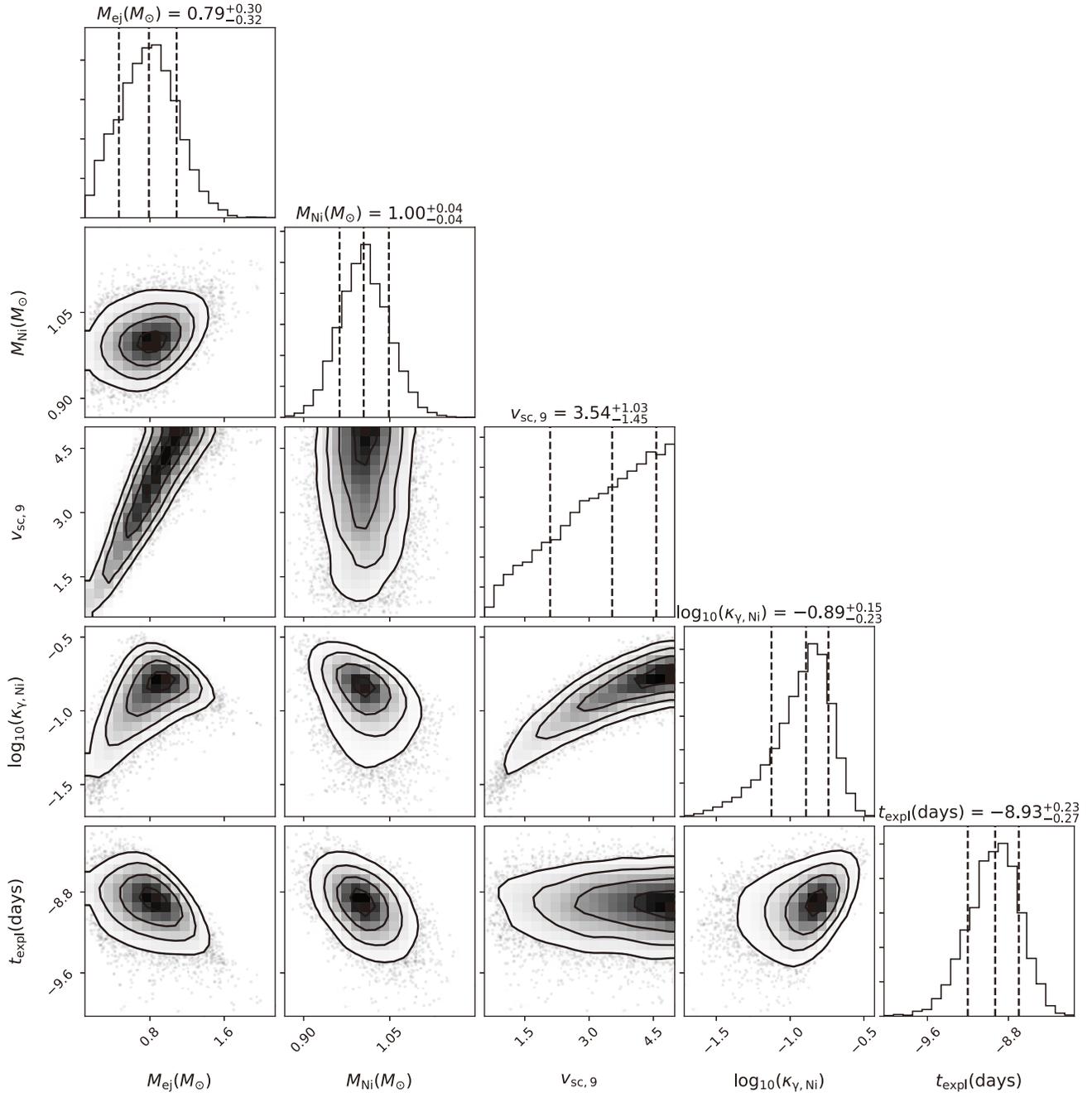}
\end{center}
\caption{The corner plot of the $^{56}$Ni model.}
\label{fig:ni_corner}
\end{figure}

The {best-fitting} LC reproduced by the $^{56}$Ni model is shown in Fig. \ref{fig:ni}
from which we can found that the $^{56}$Ni model can match the data.
The physical parameters yielding the theoretical LC are
$M_{\mathrm{ej}}=0.79_{-0.30}^{+0.32}$\,M$_\odot$,
$M_{\mathrm{Ni}}=1.00_{-0.04}^{+0.04}$\,M$_\odot$,
$v_{\mathrm{sc}}=3.54_{-1.45}^{+1.03}\times10^9$~cm s$^{-1}$,
$\kappa_{\gamma,\mathrm{Ni}}=0.13^{+0.05}_{-0.05}$~cm$^2$ g$^{-1}$,
$t_\mathrm{expl}=-8.93_{-0.27}^{+0.23}$~days;
the value of $\chi^2/\mathrm{dof}$ is $12.02/35=0.343$ (see Table \ref{tab:ni}).
Fig. \ref{fig:ni_corner} is the corner plot of the $^{56}$Ni model.

In our modeling, the value of $M_{\mathrm{ej}}$ is significantly
smaller than that of \citet{Smartt2016}
($0.79_{-0.30}^{+0.32}$\,M$_\odot$ vs 1.9\,M$_\odot$),
so that the theoretical LC is narrower than that of \citet{Smartt2016}
and can fit the data. A narrower LC require less $^{56}$Ni
($1.00_{-0.04}^{+0.04}$\,M$_\odot$ vs 1.7\,M$_\odot$) to power the
peak luminosity of the SN (the ``Arnett law", \footnote{The Arnett law
says that the peak luminosity of a SN is equal to the
input energy deposition rate at the time of the SN peak,
i.e., $L_{\rm SN,peak}=L_{\rm input}(t=t_{\rm peak})$.} \citealt{Arn1979,Arn1982}).
The lower limit of the scale velocity we inferred is approximately equal to
the value derived by \citet{Smartt2016}.

Although the theoretical LC can match the data, the {ratio of
derived $^{56}$Ni mass ($1.00_{-0.04}^{+0.04}$\,M$_\odot$) to the derived
ejecta mass ($0.79_{-0.30}^{+0.32}$\,M$_\odot$) is very large (0.86--2.12) while
the reasonable upper limit of $M_{\mathrm{Ni}}/M_{\mathrm{ej}}$ is thought to be
$\sim 0.2$ \citep{Ume2008}}, indicating that the $^{56}$Ni model is
disfavored and alternative models must be employed.

\subsection{The Magnetar Model}

The parameters of the magnetar model \citep{Kas2010,Woos2010,Inse2013,Wang2015a,Wang2016}
are {$M_{\mathrm{ej}}$, the magnetic strength of the magnetar $B_p$,
the initial rotational period of the magnetar $P_0$, $v_{\mathrm{sc}}$,
the gamma-ray opacity of magnetar photons $\kappa_{\gamma,\mathrm{mag}}$,
and $t_\mathrm{expl}$}.

\begin{figure}[tbph]
\begin{center}
\includegraphics[width=0.45\textwidth,angle=0]{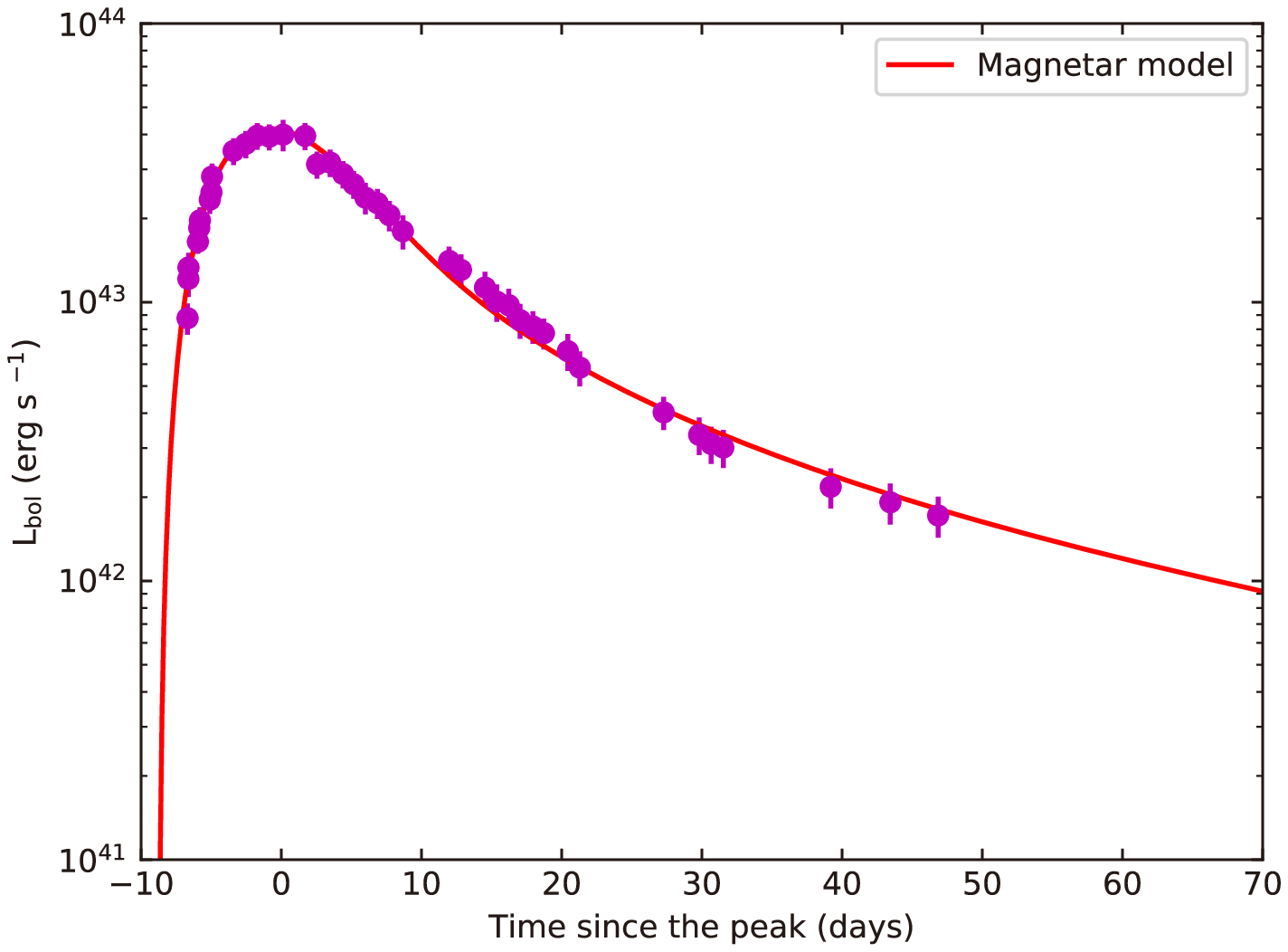}
\end{center}
\caption{The bolometric LC reproduced by the magnetar model.
Data are taken from \citet{Smartt2016}.
The abscissa represents time since the peak in the rest frame.}
\label{fig:mag}
\end{figure}

\begin{figure}[tbph]
\begin{center}
\includegraphics[width=1.0\textwidth,angle=0]{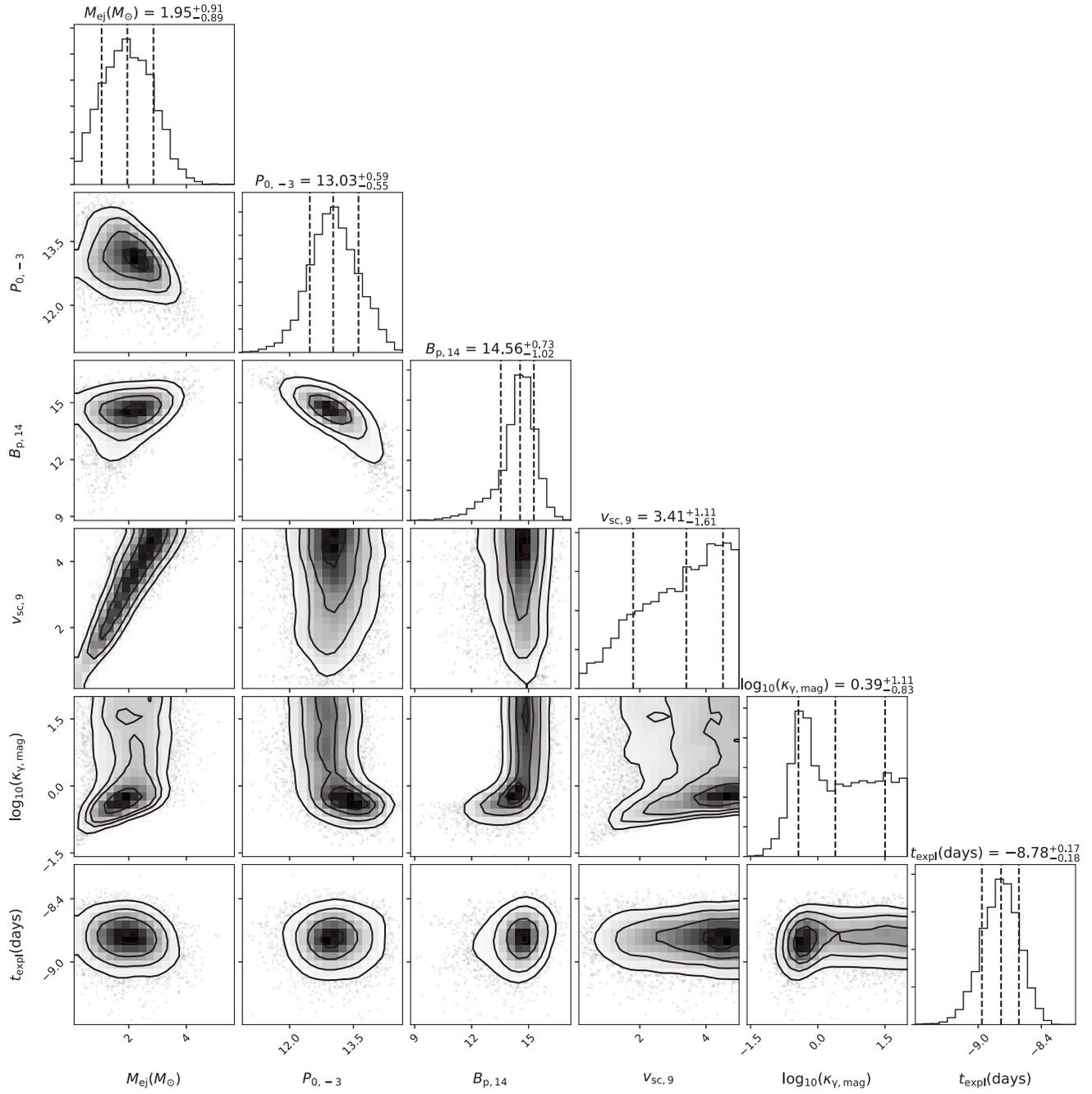}
\end{center}
\caption{The corner plot of the magnetar model.}
\label{fig:mag_corner}
\end{figure}

The theoretical LC reproduced by the magnetar model is shown in Fig. \ref{fig:mag}.
The free parameters of the magnetar model are
$M_{\mathrm{ej}}=1.95_{-0.89}^{+0.91}$\,M$_\odot$,
$B_p = 14.56_{-1.02}^{+0.73}\times10^{14}$\,G,
$P_0 = 13.03_{-0.55}^{+0.59}$\,ms,
$v_{\mathrm{sc}}=3.41_{-1.61}^{+1.11}\times10^9$~cm s$^{-1}$,
$\kappa_{\gamma,\mathrm{mag}}=2.47^{+29.36}_{-2.10}$~cm$^2$ g$^{-1}$,
$t_\mathrm{expl}=-8.78_{-0.18}^{+0.17}$~days; the value of $\chi^2/\mathrm{dof}$
is $21.15/34=0.622$ (see Table \ref{fig:mag}).
Fig. \ref{fig:mag_corner} is the corner plot of the magnetar model.


\subsection{The CSI Model}

It is widely accepted that the interaction between the SN ejecta and
the dense He-rich CSM (winds or shells) surrounding the progenitors
would provide the SNe Ibn at least a fraction of energy to power their LCs.
Therefore, the CSI model \citep{Che1982,Che1994,Chu1994,Cha2012,Gin2012,Liu2018}
taking the ejecta-CSM interaction into account is promising model to
account for the LC of PS15dpn.

The CSI models divide the SN ejecta into the inner parts \citep{Che1982}
whose density profile can be described by $\rho_{\mathrm{ej}} \propto r^{-\delta}$
and the outer part {whose density profile} can be described by
$\rho_{\mathrm{ej}} \propto r^{-n}$. We assume that the density profile of CSM can
{also} be descried by a power-law, $\rho_{\mathrm{CSM}} \propto r^{-s}$,
{where} $s=2$ corresponding to winds and $s=0$ corresponding to CSM shell.

\begin{figure}[tbph]
\begin{center}
\includegraphics[width=0.45\textwidth,angle=0]{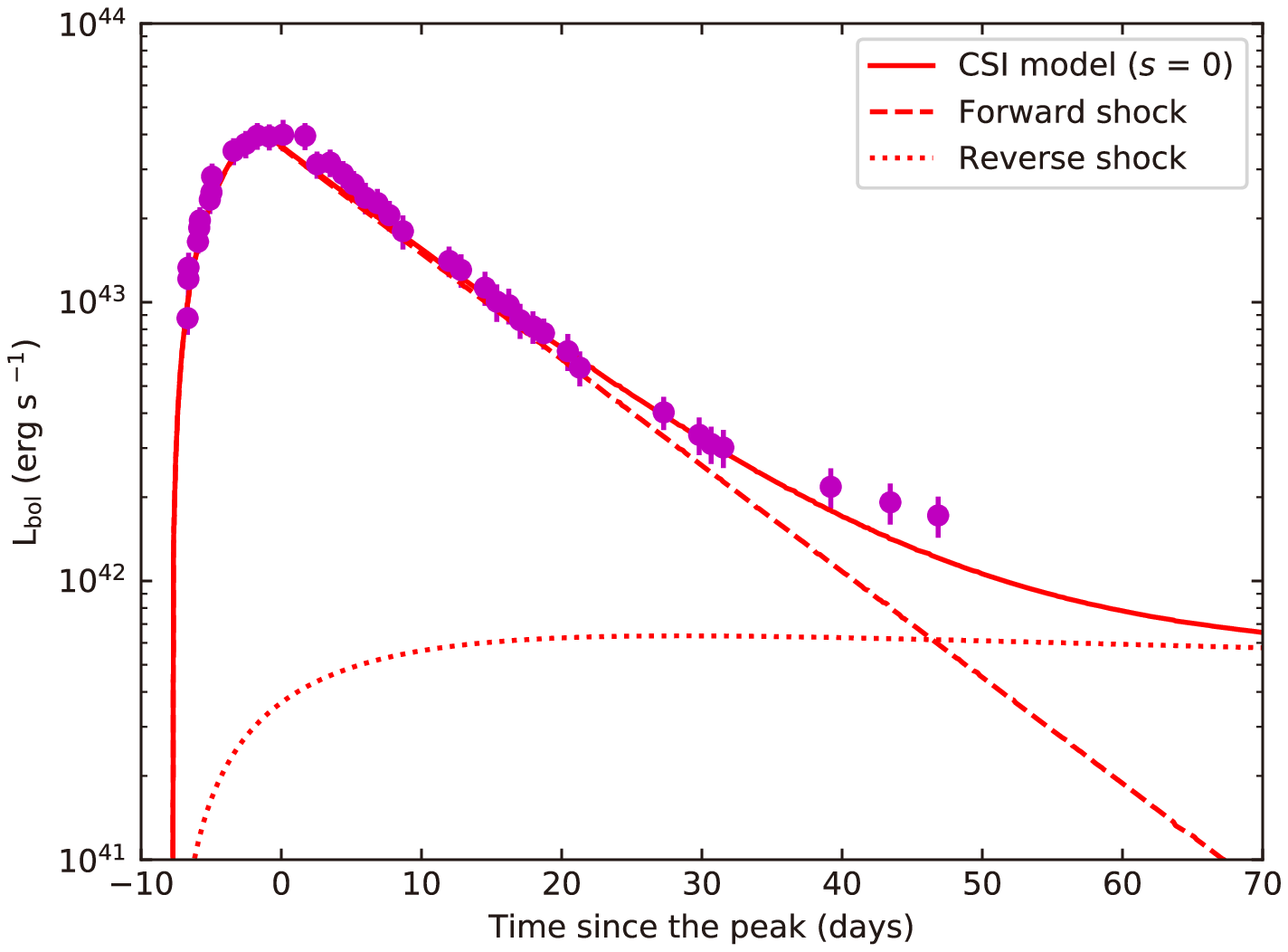}
\includegraphics[width=0.45\textwidth,angle=0]{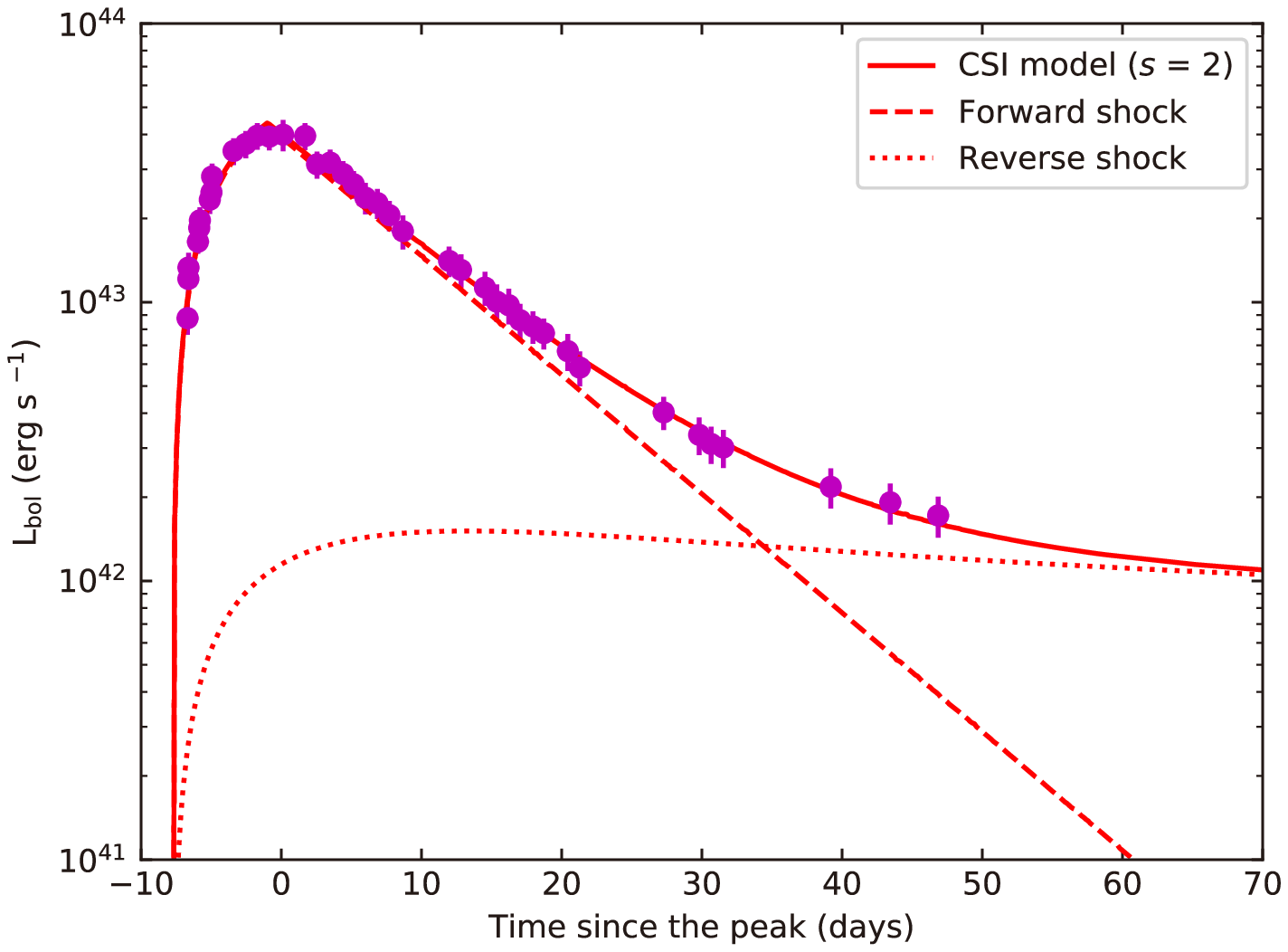}
\end{center}
\caption{The bolometric LCs reproduced by the CSI model ($s$ = 0, 2).
The LCs powered by the forward shocks and reverse shocks
are also plotted using different lines.
Data are taken from \citet{Smartt2016}.
The abscissa represents time since the peak in the rest frame.}
\label{fig:csi}
\end{figure}

Letting $\delta=1$ and $n=10$, the {free} parameters of the CSI model we adopt
are the energy of the SN ($E_{\mathrm{SN}}$), the ejecta mass ($M_{\mathrm{ej}}$),
the CSM mass ($M_{\mathrm{CSM}}$), the density of the innermost part of the CSM
($\rho_{\mathrm{CSM,in}}$), the radius of the innermost part of the CSM
($R_{\mathrm{CSM,in}}$), the efficiency of conversion from the kinetic energy
to radiation ($\epsilon$), the dimensionless position parameter ($x_{\mathrm{0}}$),
\footnote{$x=r(t)/R(t)$, the regions $x<x_{\mathrm{0}}$ and $x>x_{\mathrm{0}}$
are the inner part ($\rho_{\mathrm{ej}} \propto r^{-\delta}$) and {the} outer part
($\rho_{\mathrm{ej}} \propto r^{-n}$), respectively.} and $t_\mathrm{expl}$.
The LCs reproduced by the CSI model{s} can be found in Fig. \ref{fig:csi}
and the best-fit parameters can be found in Table \ref{tab:csi}.

\begin{figure}[tbph]
\begin{center}
\includegraphics[width=1.0\textwidth,angle=0]{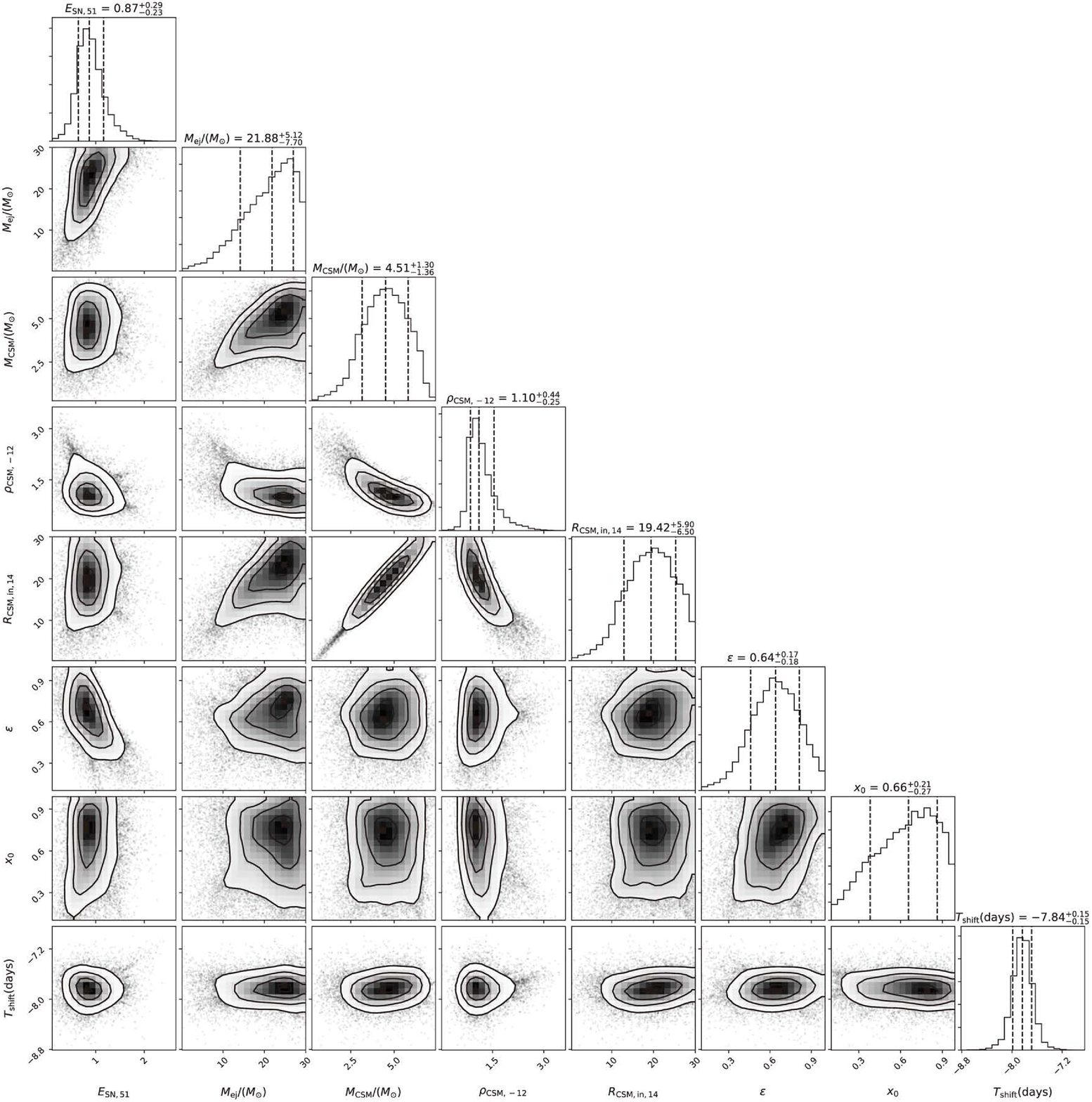}
\end{center}
\caption{The corner plot of the CSI model ($s$ = 0).}
\label{fig:csm_corner_0}
\end{figure}

For the shell ($s=0$) CSI model,
The parameters are
$E_{\mathrm{SN}}=0.87_{-0.22}^{+0.29}\times10^{51}$~erg,
$M_{\mathrm{ej}}=21.88_{-7.70}^{+5.12}$\,M$_\odot$,
$M_{\mathrm{CSM}}=4.51_{-1.36}^{+1.30}$\,M$_\odot$.
$\rho_{\mathrm{CSM,in}}=1.10_{-0.25}^{+0.44}\times10^{-12}$g~cm$^{-3}$,
$R_{\mathrm{CSM,in}}=19.42_{-6.50}^{+5.90}\times10^{14}$~cm,
$\epsilon=0.64_{-0.18}^{+0.17}$,
$x_{\mathrm{0}}=0.66_{-0.27}^{+0.21}$,
$t_\mathrm{expl}=-7.84_{-0.15}^{+0.15}$ days;
the value of $\chi^2/\mathrm{dof}$ is $29.20/32=0.912$.
Fig. \ref{fig:csm_corner_0} is the corner plot of the shell CSI model.

\begin{figure}[tbph]
\begin{center}
\includegraphics[width=1.0\textwidth,angle=0]{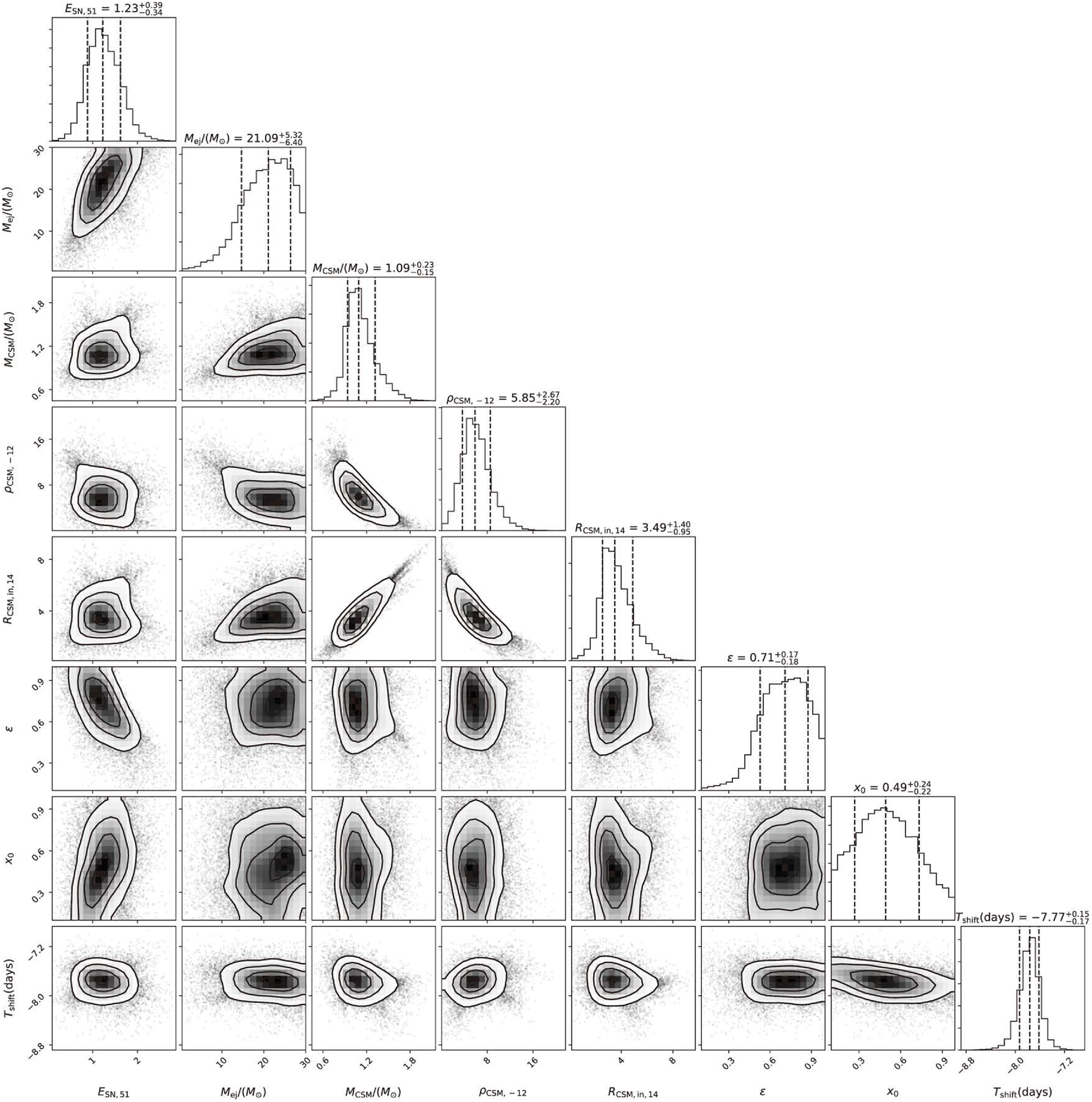}
\end{center}
\caption{The corner plot of the CSI model ($s$ = 2).}
\label{fig:csm_corner_2}
\end{figure}

For the wind ($s=2$) CSI model,
The parameters
$E_{\mathrm{SN}}=1.23_{-0.34}^{+0.39}\times10^{51}$~erg,
$M_{\mathrm{ej}}=21.09_{-6.40}^{+5.32}$\,M$_\odot$,
$M_{\mathrm{CSM}}=1.09_{-0.15}^{+0.23}$\,M$_\odot$.
$\rho_{\mathrm{CSM,in}}=5.85_{-2.20}^{+2.67}\times10^{-12}$g cm$^{-3}$,
$R_{\mathrm{CSM,in}}=3.49_{-0.95}^{+1.40}\times10^{14}$~cm,
$\epsilon=0.71_{-0.18}^{+0.17}$,
$x_{\mathrm{0}}=0.49_{-0.22}^{+0.24}$,
$t_\mathrm{expl}=-7.77_{-0.17}^{+0.15}$ days;
the value of $\chi^2/\mathrm{dof}$ is $13.58/32=0.424$.
Fig. \ref{fig:csm_corner_2} is the corner plot of the wind CSI model.
The {derived} wind mass-loss rate is $\sim$14.20 \,M$_\odot$\,yr$^{-1}$.
\footnote{The velocity of the wind ($v_{\rm w}$) of a Wolf-Rayet star
is $\sim$ 1000\,km\,s$^{-1}$, the equation calculating the mass-loss rate is
$\dot M = 4 \pi v_{\rm w} q$ ($q = \rho_{\rm CSM,in} R_{\rm CSM,in}^{2}$),
the values of $\rho_{\rm CSM,in}$ and $R_{\rm CSM,in}$ are listed in
Table \ref{tab:csi}.}

The inferred ejecta mass is {$21.88_{-7.70}^{+5.12}$}\,M$_\odot$
or {$21.09_{-6.40}^{+5.32}$\,M$_\odot$} which {are} favored by the massive
Wolf-Rayet progenitor scenario since the masses of very massive Wolf-Rayet stars
are believed to be about 25\,M$_\odot$ if their metalicity ($Z$) is equal
to that of {the Sun} ($Z_\odot$) \citep{Crow2007}.
\footnote{{The most important factors determining} the mass of the
single star just before its explosion is the Zero Age Main Sequence (ZAMS) mass
and the mass-loss rate which is dominated by line-driven wind {and} proportional to
$Z^{0.69}$ \citep{Vink2001}. {Larger metalicity results in larger mass-loss
rate and a lower mass when the progenitor explodes.} Moreover, the pre-SN eruptions
and the mass transfer in a binary system would yield a wider range of the mass of the
aged massive star.}

\subsection{The CSI Plus $^{56}$Ni Model}	

Although the CSI model can fit the LC of PS15dpn, it neglects the contribution
from $^{56}$Ni. A core collapse SN would yield a moderate amount of $^{56}$Ni.
\citet{Nom2013} {showed} that an energetic SN explosion can synthesize
$\lesssim 0.2$\,M$_\odot$ of $^{56}$Ni. Therefore, it would be more reasonable
if the luminosity from $^{56}$Ni decay is also taken into account. In other words,
the CSI plus $^{56}$Ni model \citep{Cha2012} should be adopted to fit the
bolometric LC of PS15dpn.

\begin{figure}[tbph]
\begin{center}
\includegraphics[width=0.45\textwidth,angle=0]{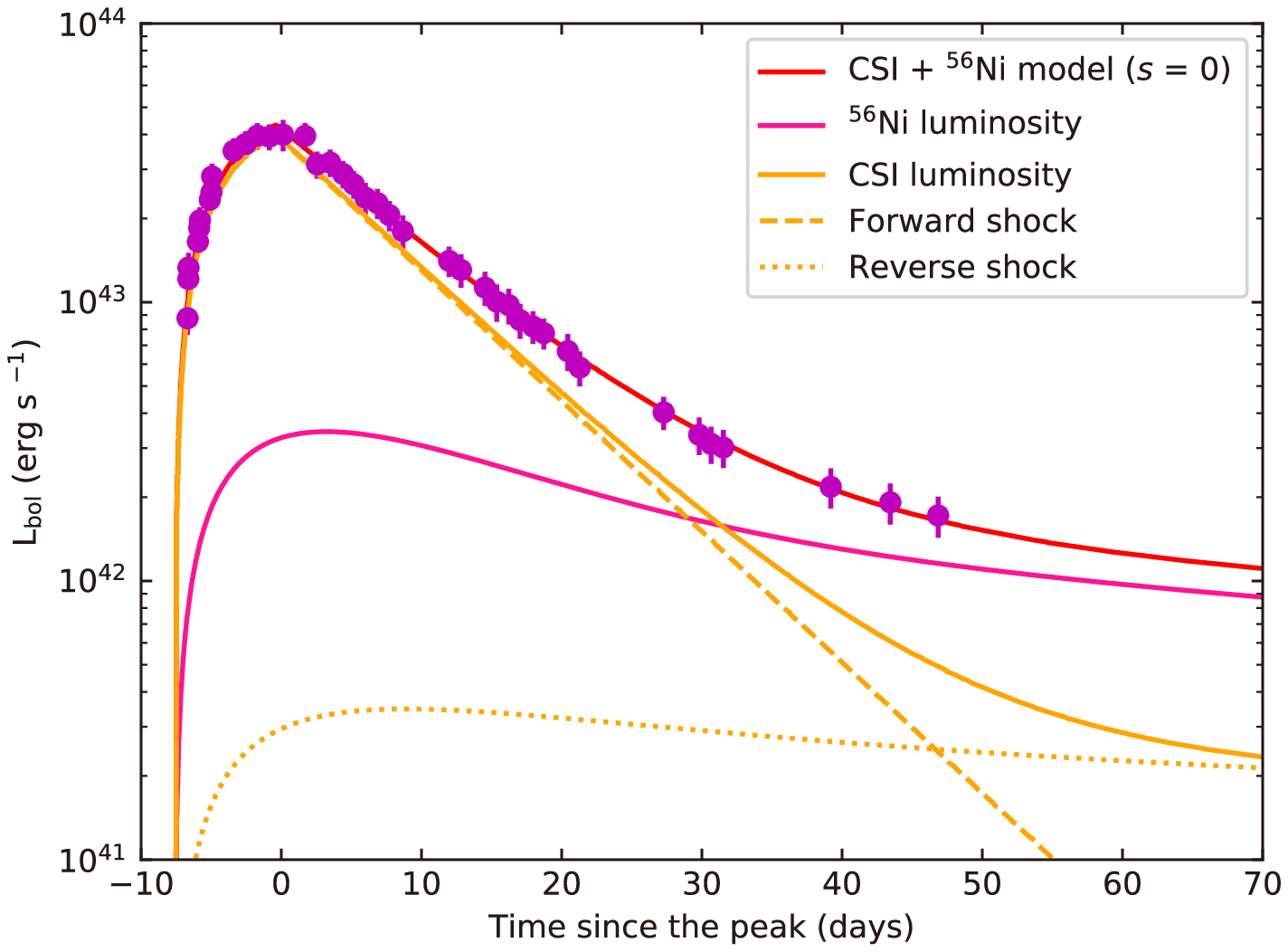}
\includegraphics[width=0.45\textwidth,angle=0]{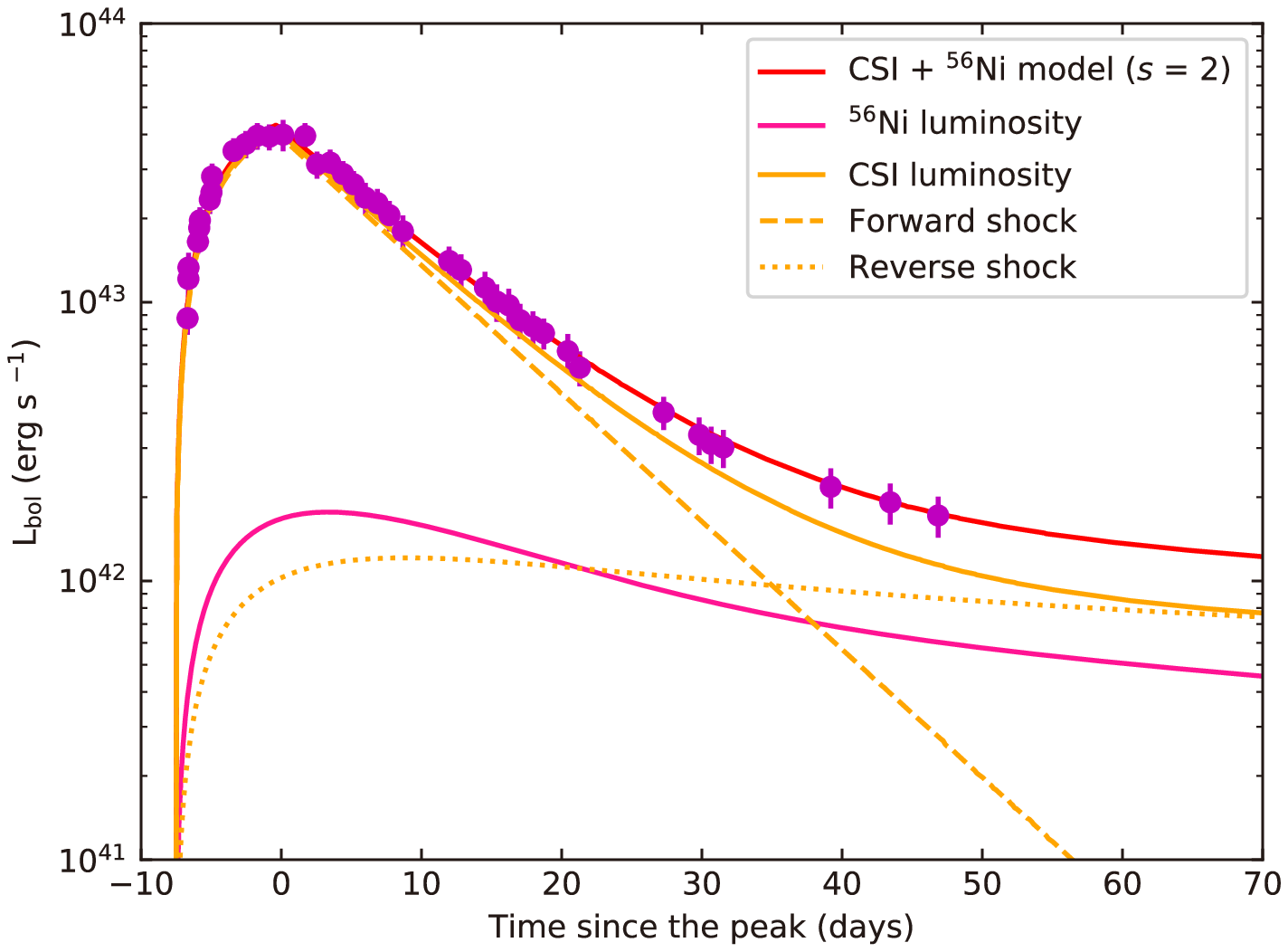}
\end{center}
\caption{The bolometric LCs reproduced by the CSI plus $^{56}$Ni model
($s$ = 0, 2). The LCs powered by the forward shocks, reverse shocks, as well as
$^{56}$Ni decay are also plotted using different lines.
Data are taken from \citet{Smartt2016}.
The abscissa represents time since the peak in the rest frame.}
\label{fig:csi+ni}
\end{figure}

Comparing with the CSI model, two additional {free} parameters
($M_{\mathrm{Ni}}$ and $\kappa_{\gamma,\mathrm{Ni}}$) must be added to
construct the CSI plus $^{56}$Ni model. The LCs reproduced by the CSI plus
$^{56}$Ni model can be found in Fig. \ref{fig:csi+ni} and the best-fit
parameters are {listed} in Table \ref{tab:csi+ni}.

\begin{figure}[tbph]
\begin{center}
\includegraphics[width=1.0\textwidth,angle=0]{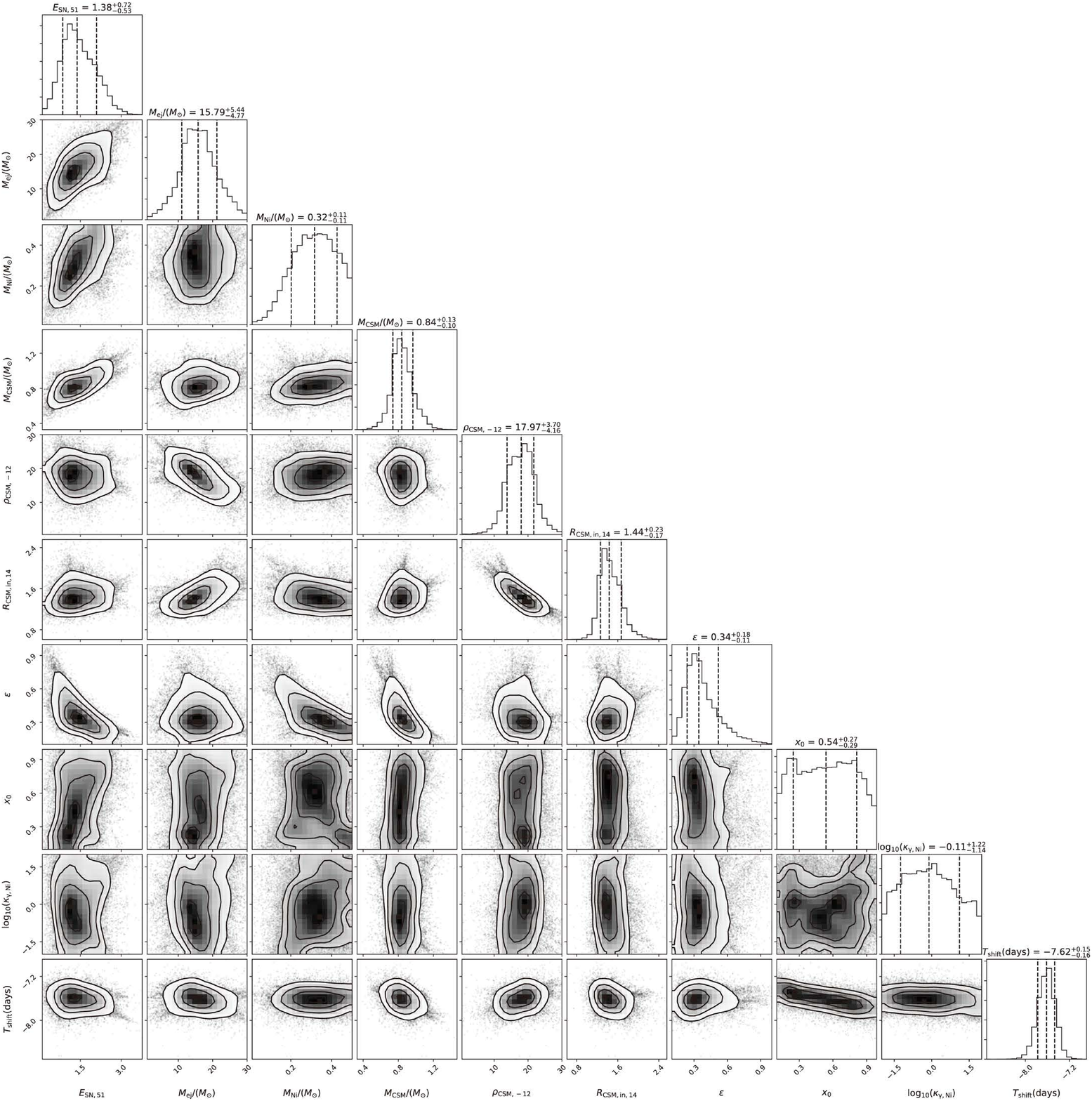}
\end{center}
\caption{The corner plot of the CSI plus $^{56}$Ni model ($s$ = 0).}
\label{fig:csm+ni_corner_0}
\end{figure}

For the shell ($s=0$) CSI plus $^{56}$Ni model,
the parameters are
$E_{\mathrm{SN}}=1.38_{-0.53}^{+0.72}\times10^{51}$~erg,
$M_{\mathrm{ej}}=15.79_{-4.77}^{+5.44}$\,M$_\odot$,
$M_{\mathrm{CSM}}=0.84_{-0.10}^{+0.13}$\,M$_\odot$,
$M_{\mathrm{Ni}}=0.32_{-0.11}^{+0.11}$\,M$_\odot$,
$\rho_{\mathrm{CSM,in}}=17.97_{-4.16}^{+3.70}\times10^{-12}$g cm$^{-3}$,
$R_{\mathrm{CSM,in}}=1.44_{-0.17}^{+0.23}\times10^{14}$~cm,
$\epsilon=0.34_{-0.11}^{+0.18}$,
$x_{\mathrm{0}}=0.54_{-0.29}^{+0.27}$,
$\kappa_{\gamma,\mathrm{Ni}}=0.78^{+12.19}_{-0.72}$~cm$^2$ g$^{-1}$,
$t_\mathrm{expl}=-7.62_{-0.16}^{+0.15}$ days;
the value of $\chi^2/\mathrm{dof}$ is $11.14/30=0.371$.
Fig. \ref{fig:csm+ni_corner_0} is the corner plot of the
shell CSI plus $^{56}$Ni model.

\begin{figure}[tbph]
\begin{center}
\includegraphics[width=1.0\textwidth,angle=0]{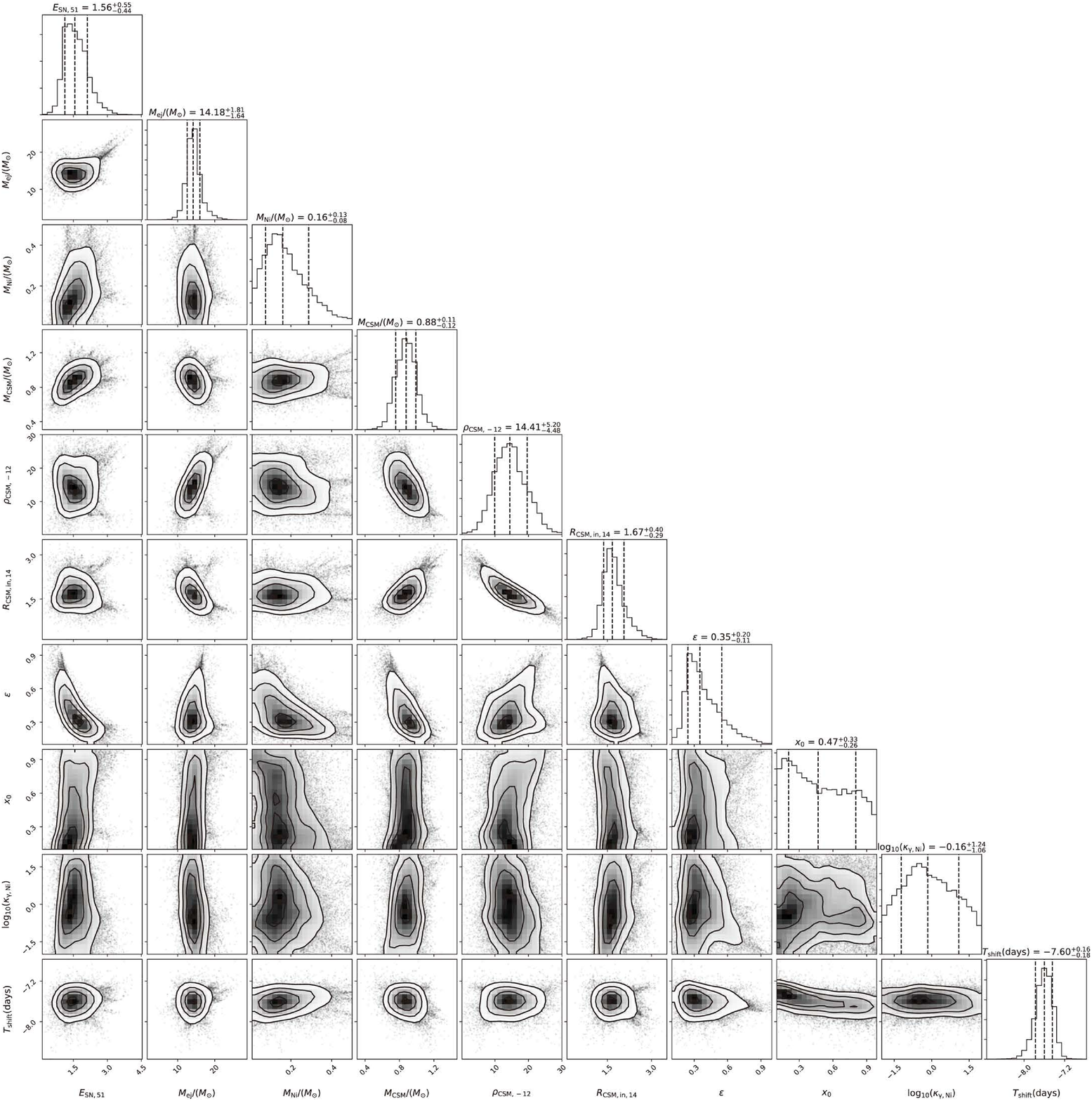}
\end{center}
\caption{The corner plot of the CSI plus $^{56}$Ni model ($s$ = 2).}
\label{fig:csm+ni_corner_2}
\end{figure}

For the wind ($s=2$) CSI plus $^{56}$Ni model,
the parameters are
$E_{\mathrm{SN}}=1.56_{-0.44}^{+0.55}\times10^{51}$~erg,
$M_{\mathrm{ej}}=14.18_{-1.64}^{+1.81}$\,M$_\odot$,
$M_{\mathrm{CSM}}=0.88_{-0.12}^{+0.11}$\,M$_\odot$,
$M_{\mathrm{Ni}}=0.16_{-0.08}^{+0.13}$\,M$_\odot$,
$\rho_{\mathrm{CSM,in}}=14.41_{-4.48}^{+5.20}\times10^{-12}$g cm$^{-3}$,
$R_{\mathrm{CSM,in}}=1.67_{-0.29}^{+0.40}\times10^{14}$~cm,
$\epsilon=0.35_{-0.11}^{+0.20}$,
$x_{\mathrm{0}}=0.47_{-0.26}^{+0.33}$,
$\kappa_{\gamma,\mathrm{Ni}}=0.69_{-0.63}^{+11.45}$~cm$^2$ g$^{-1}$,
$t_\mathrm{expl}=-7.60_{-0.18}^{+0.16}$ days;
the value of $\chi^2/\mathrm{dof}$ is $11.94/30=0.398$.
Fig. \ref{fig:csm+ni_corner_2} is the corner plot of the wind CSI plus $^{56}$Ni model.
The {derived} wind mass-loss rate is {$\sim$8.0}\,M$_\odot$\,yr$^{-1}$.

The derived masses of the the ejecta ($15.79_{-4.77}^{+5.44}$\,M$_\odot$ or
$14.18_{-1.64}^{+1.81}$\,M$_\odot$) and the CSM ($0.84_{-0.10}^{+0.13}$\,M$_\odot$
or $0.88_{-0.12}^{+0.11}$\,M$_\odot$) of these two hybrid models are reasonable
and the range of $^{56}$Ni mass ($0.32_{-0.11}^{+0.11}$\,M$_\odot$ or
$0.16_{-0.08}^{+0.13}$\,M$_\odot$) is consistent with the rough upper limit
($\lesssim 0.2$\,M$_\odot$) that can be synthesized by the explosion of a massive star.
\footnote{Systematic studies (e.g., \citealt{Lyman2016}) for SNe Ibc showed that the LCs
of a fraction of SNe Ibc that are not very luminous can be explained
by the decay of $0.2-0.6$\,M$_\odot$ of $^{56}$Ni (see, e.g., Table 5 of
\citealt{Lyman2016}). The large inferred $^{56}$Ni masses indicate that core-collapse
SNe can synthesize more than 0.2\,M$_\odot$ of $^{56}$Ni or some ordinary
SNe Ibc might be powered by hybrid energy sources. Both these two possibilities cannot
be excluded. In other words, the upper limit of the $^{56}$Ni yield of core-collapse SNe can be
larger than 0.2\,M$_\odot$.}	

\section{Discussion}
\label{sec:dis}

\subsection{Is It Necessary to Introduce A Magnetar ?}

Although the magnetar model can yield a decent fit for the LC of PS15dpn and
the parameters are reasonable, the {\it pure} magnetar model is not a promising model
accounting for the LC of PS15dpn since it neglects the contribution from the
interaction between the ejecta and the pre-existing CSM while the confirmed He {\sc i}
emission lines suggest that the ejecta-CSM interaction cannot be omitted.
The magnetar plus $^{56}$Ni model \citep{Wang2015b,Wang2016} that has also been
adopted to explain the LCs of some SNe is also disfavored since it also neglect
the contribution of CSI.

The magnetar plus CSI model or magnetar plus CSI plus $^{56}$Ni
might be promising models. From the modeling perspective, these two models are not
necessary for explaining the LC of PS15dpn since the CSI model and the CSI plus
$^{56}$Ni model can fit the LC without a putative magnetar.

It is worthwhile to note that \cite{Margutti2019} demonstrated that the LC of
AT~2018cow which is a luminous FBOT might harbor a magnetar produced
by a SN explosion or a black hole produced by a failed explosion of a blue supergiant
since the X-ray detected might be emitted by a central engine.
For PS15dpn, however, the absence of the X-ray detection prevents us from concluding
that the explosion might leave behind a magnetar or black hole.
Nevertheless, the possibility that PS15dpn leaved a magnetar or a black hole cannot be
excluded, although the CSI or the CSI plus $^{56}$Ni is sufficient to account for the
LC of PS15dpn.

\subsection{The Properties of the CSM and the Possible Pre-SN Outburst}

The values of the $\chi^2$/dof of the shell ($s=0$) CSI, the wind ($s=2$) CSI,
the shell ($s=0$) CSI plus $^{56}$Ni, and the wind ($s=2$) CSI plus $^{56}$Ni
models are 0.912, 0.424, 0.371, and 0.398. Therefore, the shell ($s=0$) CSI
plus $^{56}$Ni model is the best model since its $\chi^2$/dof value is the
smallest one.

In this model, the mass of the shell expelled from the progenitor of PS15dpn
before the SN explosion is $0.84_{-0.10}^{+0.13}$ M$_\odot$.
The radius of the innermost part of the CSM ($R_{\mathrm{CSM,in}}$) is
$\sim 1.44\times10^{14}$ cm (see Table \ref{tab:csi+ni});
assuming the speed of the shell {is} 100--1000 km s$^{-1}$,
we can {infer that the progenitor might experience an eruption
launching $0.84_{-0.10}^{+0.13}$ M$_\odot$ of material
$\sim$ 17--167} days before the SN explosion.

While the wind ($s=2$) CSI plus $^{56}$Ni model don't have the smallest
$\chi^2$/dof, it is also a promising model accounting for
the LC of PS15dpn. In this scenario, the mass-loss rate inferred
is extremely high, {$\sim 8.0$ M$_\odot$\,yr$^{-1}$}.
This values is $\sim 10^6$ times that the upper limit of line-driving stellar
wind ($\sim 10^{-5}$ M$_\odot$\,yr$^{-1}$, \citealt{Well2012}) and
significantly larger than the mass-loss rate of some SNe having
extreme pre-SN winds (e.g., iPTF13z whose mass-loss rate is
$\sim 0.1-2$ M$_\odot$\,yr$^{-1}$, \citealt{Nyh2017}).

\cite{Moriya20} proposed that the LC of the multi-peaked type IIP
SN iPTF14hls can be explained by supposing a ``super-wind" whose maximum
mass-loss rate must be $\gtrsim$ 10 $\rm M_{\odot}$ yr $^{-1}$. Although
\cite{Moriya20} pointed out that the mechanism accounting for the
production of the ``super-wind" is unclear, $\sim 8.0$ M$_\odot$\,yr$^{-1}$
of mass-loss rate and the super-wind scenario cannot be excluded.

\subsection{Comparison with other Type Ibn SNe and FBOTs}

It is interesting to compare the properties of PS15dpn with other
Type Ibn SNe and FBOTs. To perform the comparison, some phase-space figures
are needed. \cite{Margutti2019} plotted a phase-space figure showing the values
of peak luminosity and rise time of AT~2018cow and other optical transients
(including other FBOTs, SNe Ibc, SN II, as well as SLSNe), see their Fig. 1.
\citet{Fox2019} showed in their Figs. 2 and 3 three phase-space figures
demonstrating the peak luminosity, the rise time, and the decline rate of
AT~2018cow, other FBOTs, as well as SNe Ibn.

\begin{figure}[tbph]
\begin{center}
\includegraphics[width=0.45\textwidth,angle=0]{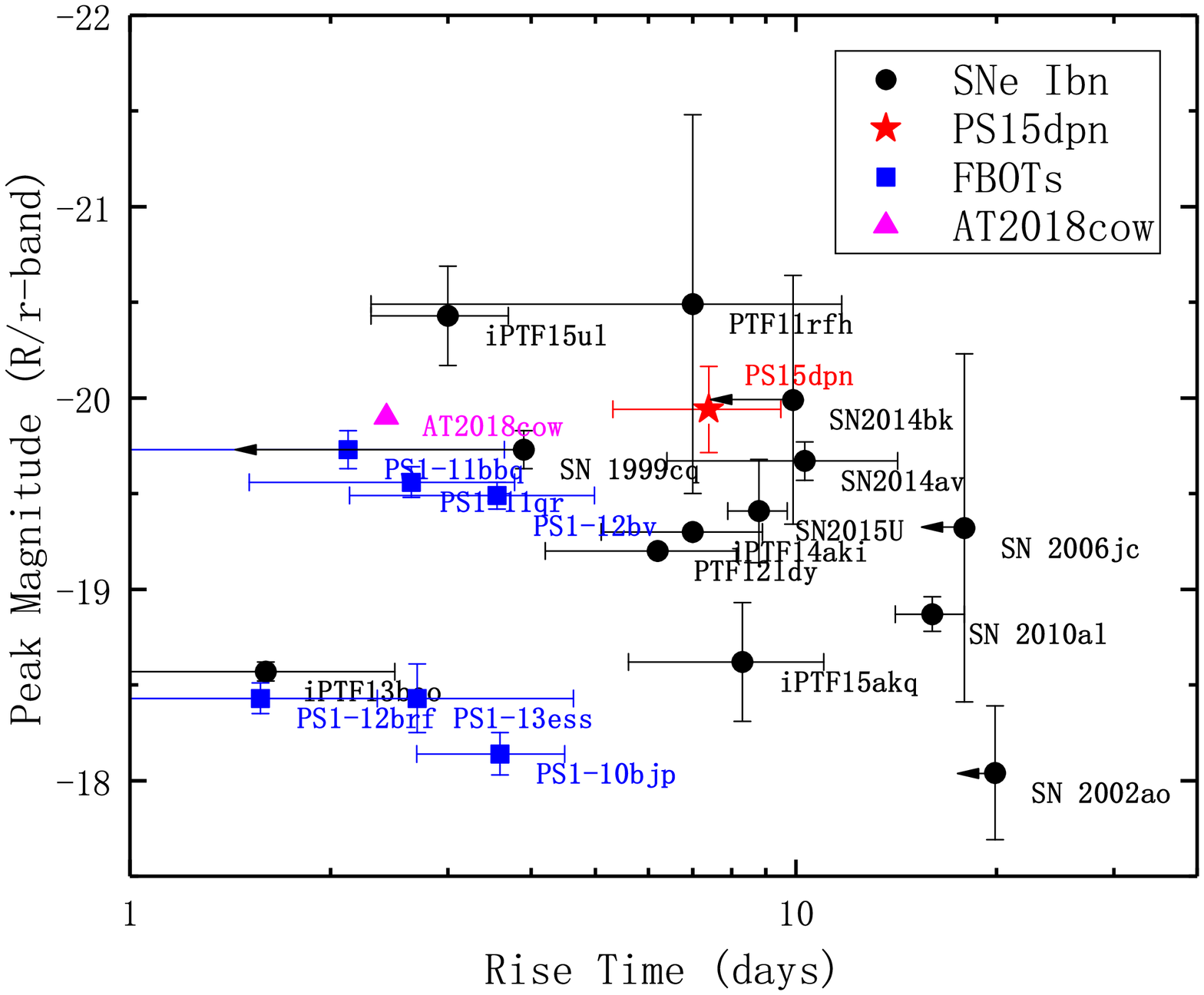}
\includegraphics[width=0.45\textwidth,angle=0]{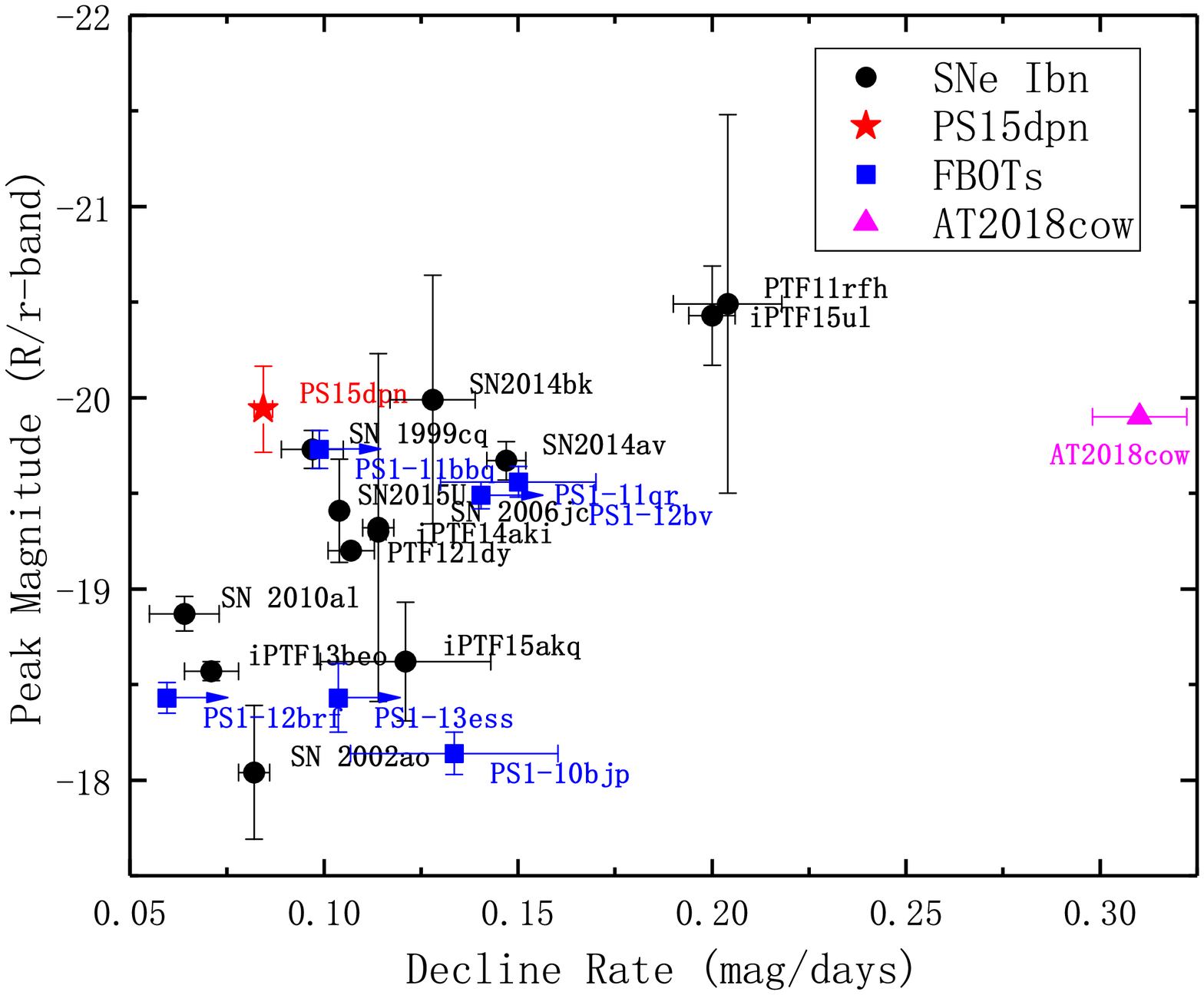}
\includegraphics[width=0.45\textwidth,angle=0]{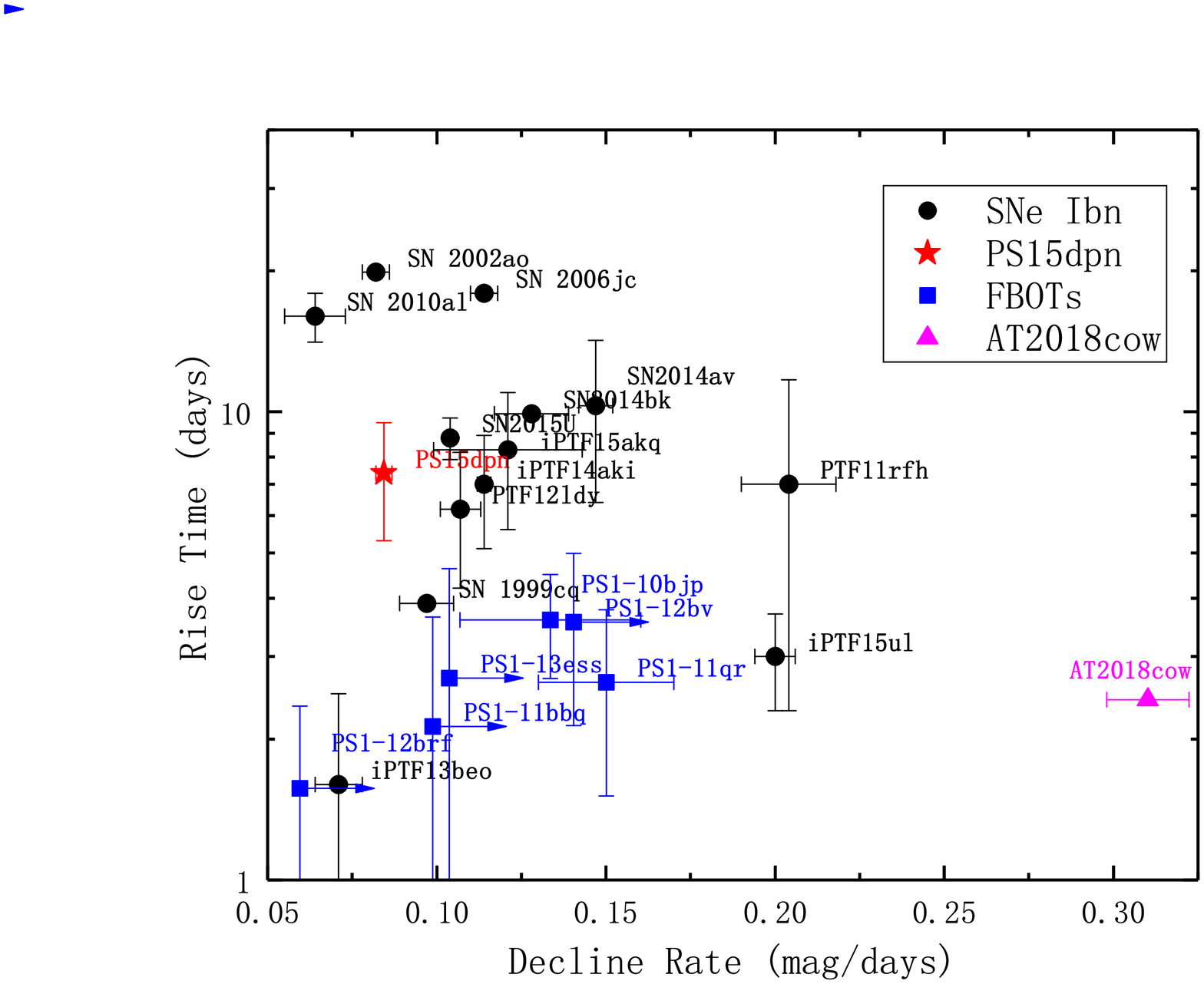}
\end{center}
\caption{Phase-space plots showing peak magnitudes versus rise time (top left), peak magnitudes versus decline rates (top right),
and rise time versus decline rates (bottom) of PS15dpn (red star) SNe Ibn (black filled circles),
AT~2018cow (magenta filled  triangle), and FBOTs discovered by Pan-STARRS1 (blue filled squares).}
\label{fig:phase-space}
\end{figure}

We also plot three $R/r$ band phase-space figures showing peak
magnitude versus rise time, peak magnitude versus decline rate, and rise
time versus decline rate of SNe Ibn (including PS15dpn) and FBOTs (including
AT~2018cow), see Figure \ref{fig:phase-space}. The values of these quantities
of SNe Ibn and FBOTs discovered by Pan-STARRS come from Table 4 of
\cite{Hoss2017} (and the references therein)\footnote{We don't include
the events whose rise time were loosely constrained, e.g., SN 2011hw
($t_{\rm rise}<334$ days); SN~2000er ($t_{\rm rise}<46.6$ days).}
and Table 4 of \cite{Drout2014}, respectively; for AT~2018cow and
PS15dpn, the values are taken from \cite{Per2019} and \cite{Smartt2016},
respectively. The value of $r$-band rise time of PS15dpn is absent and
we suppose that the rise time of $i$-band is equal to that of $r$-band.

Unlike AT~2018cow which shows extreme features in the phase-space
plots, PS15dpn is a common event in the SNe Ibn sample. While the
$R$-band peak magnitude of PS15dpn is approximately equal to that of
AT~2018cow, the decline rate of PS15dpn is comparable to the mean value
of the decline rates of the SNe Ibn sample. Furthermore,
the rise time of PS15dpn is also approximately equal to the mean value
of the rise time of SNe Ibn and significantly larger than that of AT~2018cow
and other FBOTs. These facts indicate that PS15dpn is a typical SN Ibn and
its origin might be different from that of many FBOTs.

\section{Conclusions}

\label{sec:con}

PS15dpn is a luminous rapidly rising SN Ibn discovered by Pan-STARRS1.
\citet{Smartt2016} demonstrated that PS15dpn cannot be
explained by $^{56}$Ni model. In this paper, we
investigate the possible energy sources that can power
the bolometric LC of PS15dpn.

We found that the $^{56}$Ni model is disfavored since the the
ratio of inferred $^{56}$Ni to the derived ejecta is too large, supporting
the conclusion of \citet{Smartt2016}. The magnetar model can fit the data,
but it omits the contribution from the interaction between the ejecta
and the CSM {while} the confirmed He {\sc i} emission
lines indicative of the interaction between the ejecta and the dense
He-rich CSI provide evidence that the ejecta-CSM interaction would
release a fraction of energy to power the LC of PS15dpn.
This means that the {\it pure} magnetar model is not a reasonable
model, though the possibility that the explosion of PS15dpn cannot be
excluded.

Both the CSI model and the CSI plus $^{56}$Ni model can fit the data and
the inferred ejecta mass are about 14--22\,M$_\odot$ which is consistent
with the scenario that the progenitors of SNe Ibn are very massive Wolf-Rayet
stars surrounded by the dense He-rich CSM yielded by the eruptions
{or winds} of the progenitors {\citep{Fol2007}}.

The models containing $^{56}$Ni contribution {are} more reasonable
since {the explosions of massive stars would synthesize a moderate
amount of $^{56}$Ni ($M_{\mathrm{Ni}}\lesssim 0.2$\,M$_\odot$, see, e.g.,
\citealt{Nom2013}) and the values of $\chi^2$/dof of the CSI plus $^{56}$Ni
models are smaller than that of the CSI models.}
The inferred $^{56}$Ni mass of the shell (wind) CSI plus $^{56}$Ni model is
{$0.32_{-0.11}^{+0.11}$\,M$_\odot$ ($0.16_{-0.08}^{+0.13}$}\,M$_\odot$),
consistent with the rough upper limit of the $^{56}$Ni yield that can be synthesized
by core-collapse SNe.

Based on the physical parameters derived, we can explore the mass
loss history of the progenitor of PS15dpn. We found that the progenitor
expelled a shell $\sim$ 17--167 days before the SN explosion if
the LC of PS15dpn was powered by the ejecta-shell interaction and $^{56}$Ni.
Alternatively, the progenitor might blow a ``super-wind" whose mass-loss
rate is $\sim 8.0$ M$_\odot$\,yr$^{-1}$ if the the LC of PS15dpn was powered
by the ejecta-wind interaction and $^{56}$Ni.

To date, the unique SN Ibn {having} pre-SN outburst observed
is SN~2006jc \citep{Fol2007,Pas2007} whose precursor outburst was recorded in
2014. It is worthy to search the possible precursor explosion from the archival
data of other SNe Ibn.
But it is not surprising that no precursor is discovered since the optical
display associated with the precursor outburst or super-wind of PS15dpn
{might be} too dim to be detected or no telescopes observed the position
of PS15dpn when the light emitted by the pre-SN outburst or super-wind
arrived the earth.

It can be expected that current {and} future optical sky survey facilities
would discover plenty of precursor outbursts followed by the explosions of
SNe Ibn, shedding more light on the nature of SNe Ibn and helping
investigate the mass-loss history of SNe Ibn.

\acknowledgments
We thank anonymous referees for helpful comments and
suggestions that have allowed us to improve this manuscript.
This work is supported by National Natural Science Foundation of China
(grants 11963001, 11533003, 11603006, 11673006, 11851304, and U1731239),
Guangxi Science Foundation (grants 2016GXNSFCB380005, 2016GXNSFFA380006
and 2017GXNSFFA198008, AD17129006, and 2018GXNSFGA281007), and the Bagui Young
Scholars Program (LHJ).

\clearpage

\begin{table*}[tbp]
\caption{Parameters of the $^{56}$Ni model. The uncertainties are 1$\sigma$.}
\label{tab:ni}
\begin{center}
{\scriptsize
\begin{tabular}{ccccccccccc}
\hline\hline
	&	$M_{\mathrm{ej}}$ &	$M_{\mathrm{Ni}}$ &	 $v_{\mathrm{sc}}$ 	 &$\log(\kappa_{\rm \gamma,Ni})$ &	$t_\mathrm{expl}$$\star$	& $\chi^2/\mathrm{dof}$ \\
	&	 (M$_{\odot}$) 	  &	 (M$_{\odot}$) 	  &	 ($10^9$cm s$^{-1}$) &	 (cm$^2$ g$^{-1}$) 	        &	 	 (days)	&	\\
\hline
\hline
 uniform priors & [0.1, 20.0]  &[0.1, 5.0] &	[0.1, 5.0]  & [$-$2.0, 2.0]  &  [$-$14.0, $-$6.0]   &	   \\
\hline
  	&	 $0.79_{-0.30}^{+0.32}$  	&	 $1.00_{-0.04}^{+0.04}$  	&	 $3.54_{-1.45}^{+1.03}$  	&	 $-0.89_{-0.23}^{+0.15}$  	&		 $-8.93_{-0.27}^{+0.23}$  	&	$12.02/35$\\
\hline
\hline
\end{tabular}}
\end{center}
\par
{$\star$ The value of $t_\mathrm{expl}$ is with respect to the date of the SN peak. \newline}
\end{table*}

\begin{table*}[tbp]
\caption{Parameters of the magnetar model. The uncertainties are 1$\sigma$.}
\label{tab:mag}
\begin{center}
{\scriptsize
\begin{tabular}{ccccccccccc}
\hline\hline
	&$M_{\mathrm{ej}}$ 	&$P_0$ 	&	 $B_p$   	&	 $v_{\mathrm{sc}}$ 	&	  $\log(\kappa_{\rm \gamma,mag})$ 	&	 $t_\mathrm{expl}$	&	 $\chi^2/\mathrm{dof}$ \\
	&(M$_{\odot}$)	    &(ms)	& ($10^{14}$~G) &	 ($10^9$cm s$^{-1}$) &	 (cm$^2$ g$^{-1}$) 	&	 (days)	&	\\
\hline
\hline
 uniform priors & [0.1, 20.0]  &  [0.1, 100.0] &	[0.1, 100.0]  & [0.1, 5.0]  & [$-$2.0, 2.0]  & [$-$14.0, $-$6.0]  &	          \\
\hline
 	& $1.95_{-0.89}^{+0.91}$  &	 $13.03_{-0.55}^{+0.59}$ &	$14.56_{-1.02}^{+0.73}$  & $3.41_{-1.61}^{+1.11}$  & $0.39_{-0.83}^{+1.11}$  & $-8.78_{-0.18}^{+0.17}$  &	$21.15/34$\\
\hline
\hline
\end{tabular}}
\end{center}
\par
\end{table*}


\setlength{\tabcolsep}{1mm}
\begin{table*}[tbp]
\caption{Parameters of the CSI model. The uncertainties are 1$\sigma$.}
\label{tab:csi}
\begin{center}
{\scriptsize
\begin{tabular}{cccccccccccccc}
\hline
\hline
&	$s$	&	 $E_{\mathrm{SN}}$	& $M_{\mathrm{ej}}$ & $M_{\mathrm{CSM}}$ &	 $\rho_{\mathrm{CSM,in}}$ &	$R_{\mathrm{CSM,in}}$ &	$\epsilon$ & $x_{\mathrm{0}}$	&	 $t_\mathrm{expl}$	& $\chi^2/\mathrm{dof}$ \\
&		&	 ($10^{51}$~erg) 	&	 (M$_{\odot}$) 	&	 (M$_{\odot}$) 	&	  ($10^{-12}$g cm$^{-3}$) 	&	 ($10^{14}$cm) 	&		 &			 &	 (days)	&	 \\
\hline
\hline
uniform priors&	 	&	[0.1, 5.0]	& [0.1, 30.0] & [0.1, 20.0] &	[0.01, 30.0] &	[0.01, 30.0] &	[0.1, 1.0] & [0.1, 1.0]	&	[$-$14.0, $-$6.0]	&  \\
\hline
&	0		&	 $0.87_{-0.22}^{+0.29}$  	&	 $21.88_{-7.70}^{+5.12}$  	&	$4.51_{-1.36}^{+1.30}$	&		 $1.10_{-0.25}^{+0.44}$  	 &	 $19.42_{-6.50}^{+5.90}$  	&	 $0.64_{-0.18}^{+0.17}$  	&	 $0.66_{-0.27}^{+0.21}$  	&	 $-7.84_{-0.15}^{+0.15}$  	&	$29.20/32$\\
&	2		&	 $1.23_{-0.34}^{+0.39}$  	&	 $21.09_{-6.40}^{+5.32}$  	&	$1.09_{-0.15}^{+0.23}$	&	 	 $5.85_{-2.20}^{+2.67}$  	&	 $3.49_{-0.95}^{+1.40}$  	&	 $0.71_{-0.18}^{+0.17}$  	&	 $0.49_{-0.22}^{+0.24}$  	&	 $-7.77_{-0.17}^{+0.15}$  	&	$13.58/32$\\
\hline
\hline
\end{tabular}}
\end{center}
\par
\end{table*}

\setlength{\tabcolsep}{1mm}
\begin{table*}[tbp]
\caption{Parameters of the CSI plus $^{56}$Ni model. The uncertainties are 1$\sigma$.}
\label{tab:csi+ni}
\begin{center}
{\scriptsize
\begin{tabular}{cccccccccccccc}
\hline
\hline
	&	$s$	&	 	$E_{\mathrm{SN}}$	&	 $M_{\mathrm{ej}}$	&	 $M_{\mathrm{Ni}}$ & $M_{\mathrm{CSM}}$ 	 	&	 $\rho_{\mathrm{CSM,in}}$ 	&	 $R_{\mathrm{CSM,in}}$ 	&	 $\epsilon$	 &	$x_{\mathrm{0}}$	&	 $\log(\kappa_{\rm \gamma,Ni})$ 	&	 $t_\mathrm{expl}$	&	 $\chi^2/\mathrm{dof}$ \\
	&		 	&	 ($10^{51}$~erg) 	&	 (M$_{\odot}$) 	&	 (M$_{\odot}$) 	&	 (M$_{\odot}$) 	&	 ($10^{-12}$g cm$^{-3}$) 	&	 ($10^{14}$cm) 	&		 &		&	 (cm$^2$ g$^{-1}$) 	 &	 (days)	&	\\
\hline
\hline
uniform priors&	 	&	[0.1, 5.0]	& [0.1, 30.0] & [0.01, 0.5] & [0.1, 20.0] &	[0.01, 30.0] &	[0.01, 30.0] &	[0.1, 1.0] & [0.1, 1.0]	 & [$-$2.0, 2.0] &	 [$-$14.0, $-$6.0]	&  \\
\hline
&	0	&	 $1.38_{-0.53}^{+0.72}$  	&	 $15.79_{-4.77}^{+5.44}$  &	$0.32_{-0.11}^{+0.11}$   	&	 $0.84_{-0.10}^{+0.13}$   		 &	 $17.97_{-4.16}^{+3.70}$  	&	 $1.44_{-0.17}^{+0.23}$  	&	 $0.34_{-0.11}^{+0.18}$  	 &	 $0.54_{-0.29}^{+0.27}$  	&	$-0.11_{-1.14}^{+1.22}$  	 &	 $-7.62_{-0.16}^{+0.15}$  	&	$11.14/30$\\
&	2		&	 $1.56_{-0.44}^{+0.55}$  	&	 $14.18_{-1.64}^{+1.81}$  	&	$0.16_{-0.08}^{+0.13}$  	&	 $0.88_{-0.12}^{+0.11}$   	 &	 $14.41_{-4.48}^{+5.20}$  	 &	 $1.67_{-0.29}^{+0.40}$  	&	 $0.35_{-0.11}^{+0.20}$  	&	 $0.47_{-0.26}^{+0.33}$  	&	$-0.16_{-1.06}^{+1.24}$  	 &	 $-7.60_{-0.18}^{+0.16}$  	&	 $11.94/30$\\
\hline
\hline
\end{tabular}}
\end{center}
\par
\end{table*}

\clearpage

\clearpage

\clearpage

\clearpage

\clearpage

\clearpage

\clearpage

\clearpage

\clearpage

\clearpage

\clearpage

\end{document}